
\input phyzzx

\nopagenumbers
\line{\hfil CU-TP-618}
\line{\hfil hep-th/9312074}
\vglue .4in
\centerline {\twelvebf  Fermions in a Global Monopole Background }
\vskip .3in
\centerline{\it Hai Ren }
\vskip .1in
\centerline{Physics Department}
\centerline{Columbia University }
\centerline{New York, New York 10027}
\vskip .4in
\baselineskip=20pt
\overfullrule=0pt
\centerline {\bf Abstract}
\medskip

	We study the scattering of fermions in a ``global monopole''
background metric. This is the four-dimensional analogue of the
scattering on a cone in three dimensions. The scattering amplitude
is exactly obtained. We then study massless fermion-dyon systems in
such a background metric. The density of the $S$-wave fermion
condensate is found to be given by a constant times the flat space
value of Callan and Rubakov.

\vskip 1.1in
\noindent\footnote{}{\twelvepoint This work was supported in part by
the US Department of Energy }

\vfill
\eject

\baselineskip=19pt
\pagenumbers
\pageno=1

\def\refmark#1{[#1]}
\def\ga{\alpha}
\def\gb{\beta}
\def\gd{\delta}
\def\gD{\Delta}

\def\w{\omega}
\def\ge{\epsilon}
\def\gl{\lambda}
\def\gg{\gamma}
\def\bgg{{\gg\kern -0.52em \gg}}
\def\gs{\sigma}
\def\bgs{{\sigma\kern -0.54em \sigma}}
\def\bgt{{\tau\kern -0.60em \tau}}
\def\gG{\Gamma}
\def\p{\partial}
\def\gp{\varphi}
\def\gt{\theta}
\def\bn{{\bf n}}
\def\br{{\bf r}}
\def\bp{{\bf p}}
\def\tgg{{\tilde \gg}}
\def\sqr#1#2{{\vcenter{\hrule height.#2pt
        \hbox{\vrule width.#2pt height#1pt \kern#1pt
                \vrule width.#2pt}
        \hrule height.#2pt}}}
\def\square{\mathchoice\sqr64\sqr64\sqr{2.1}3\sqr{1.5}3}

{\it 1. Scattering of Fermions}

	We study the fermion scattering in a ``global monopole''
background metric (whose explicit form we will give below). This is one
of the few examples where the scattering amplitude can be exactly
obtained. It is the four-dimensional analogue of the scattering on a cone
studied in \Ref\Deser{S.Deser and R.Jackiw, Comm.Math.Phys. {\bf 118},
495 (1988); P.Gerbert and R.Jackiw, {\it ibid.} {\bf 124}, 229 (1989).}.
The scattering of scalar particles in such a background is studied in
\Ref\Mazur{P.Mazur and J.Papavassiliou, Phys.Rev.D{\bf 44}, 1317 (1991).}.

	First, we consider the general spherically symmetric metric
$$ ds^2 = \gb^2(r)dt^2 -{1\over \ga^2(r)}dr^2 - R^2(r)(d\theta^2 +
\sin^2\theta d\varphi^2)
	\eqn\general   $$
We choose the {\it vierbein} such that the fermion wave functions
will be single-valued
$$ e^\mu_a = \left(\matrix{{1\over\gb}&0&0&0\cr
0&\ga\sin\theta\cos\gp&{1\over R}\cos\theta
\cos\gp&-{1\over R\sin\theta}\sin\gp\cr
0&\ga\sin\theta\sin\gp&{1\over R}\cos\theta
\sin\gp&{1\over R\sin\theta}\cos\gp\cr
0&\ga\cos\theta&-{1\over R}\sin\theta&0\cr } \right )
	\eqn\siii   $$
After standard manipulations, we arrive at the Dirac equation
$$ i\left[{1\over \gb}\gg^0\p_t + i\bgg\cdot\bp^{(\ga,R)} -
{1-\ga R'\over R}\bgg\cdot
\bn + \ga{\gb'\over 2\gb}\bgg\cdot\bn \right]\psi = M\psi
	\eqn\sxi    $$
where
$$i\bgg\cdot\bp^{(\ga,R)} \equiv
\bgg\cdot\bn\ \ga(r)\p_r
+ \bgg\cdot\p_\gt\bn\ {1\over R(r)}\p_\gt +
\bgg\cdot(\bn\times\p_\gt\bn)\ {1\over R(r)\sin\gt}\p_\gp
	\eqn\sxii    $$
$\bn \equiv \hat\br$, and a prime denotes differentiation with
respect to $r$.

	Next we specialize to the global monopole background metric
$$ ds^2 = dt^2 - {1\over \ga^2}dr^2 - r^2(d\gt^2 + \sin^2\gt d\gp^2)
	\eqn\ai  $$
where $\ga \le 1$ is a constant. It will be convenient to use the
representation of the flat space $\gg$-matrices
$$ \gg^0 = \left(\matrix{1&0\cr0&-1}\right)\ , \qquad
\bgg = \left(\matrix{0&\bgs\cr-\bgs&0}\right)
	\eqn\aiii  $$
A complete set of solutions to the Dirac equation
are of the form $\psi^{(1)}_{jm}e^{-iEt}$
and $\psi^{(2)}_{jm}e^{-iEt}$, with
$$ \psi^{(1)}_{jm} = \left[\matrix{f(r)\phi^{(1)}_{jm} \cr
g(r)\phi^{(2)}_{jm} }\right]\ , \qquad
\psi^{(2)}_{jm} = \left[\matrix{f(r)\phi^{(2)}_{jm} \cr
g(r)\phi^{(1)}_{jm} }\right]
	\eqn\aiv   $$
where $\phi^{(1)}_{jm}$ and $\phi^{(2)}_{jm}$
are spinor spherical harmonics. For $\psi^{(1)}_{jm}$, the Dirac
equation becomes
$$\eqalign{ (E-M)f &= i\left(\p_r + {1+\nu\over r}\right)g  \cr
 (E+M)g &= i\left(\p_r + {1-\nu\over r}\right)f  }
	\eqn\avi  $$
where $\nu\equiv (j+1/2)/\ga$, and we have redefined $E$ and $M$ by letting
$E/\ga \rightarrow E , M/\ga \rightarrow M $.
The solution is given in terms of Bessel functions
$$f={ik\over E-M}{1\over\sqrt{kr}}J_{\nu-1/2}(kr) \ ,\qquad
g={1\over\sqrt{kr}}J_{\nu+1/2}(kr)
	\eqn\avii  $$
where $k\equiv \sqrt{E^2-M^2}$. Similarly for $\psi^{(2)}_{jm}$,
we have
$$f= {1\over\sqrt{kr}}J_{\nu+1/2}(kr) \ ,\qquad
g= {ik\over E+M}{1\over\sqrt{kr}}J_{\nu-1/2}(kr)
	\eqn\aix  $$

	We can now construct the scattering solution. The incident
wave, with helicity $+1$, has four components. However,
when the upper two components are matched, the lower two components
will automatically be matched. Also the overall normalization factor
will cancel out at the end of the calculation.
We may therefore only consider the upper two components of
the incident wave
$$\psi_{\rm in} = e^{-ikz}\left(\matrix{0\cr1}\right)
	\eqn\axi  $$
whose large $r$ behavior is
$$\eqalign{e^{-ikz}\left(\matrix{0\cr1}\right) &\longrightarrow
-{e^{ikr}\over ikr}\left({\pi\over 2}\right)^{1/2}\sum^\infty_{j=1/2}
\sqrt{2j+1}\ (-1)^{j+1/2}\left(\phi^{(1)}_{jm} - \phi^{(2)}_{jm}\right)
\cr
&\qquad -{e^{-ikr}\over ikr}\left({\pi\over 2}\right)^{1/2}
\sum^\infty_{j=1/2}
\sqrt{2j+1}\left(\phi^{(1)}_{jm} + \phi^{(2)}_{jm}\right) }
	\eqn\axiii  $$
where $m\equiv -1/2$.  It follows that the
scattering solution is given by
$$ \psi = \pi{E-M\over k}\sum_{j=1/2}^\infty e^{-i\pi\nu/2}\sqrt{2j+1}\left(
\psi^{(1)}_{jm} + {k\over E-M}\psi^{(2)}_{jm}\right)
	\eqn\axiv  $$
The outgoing wave, defined by $\psi = \psi_{\rm in} + \psi_{\rm out}$,
is given for large $r$ by
$$\psi_{\rm out} = {e^{ikr}\over ikr}\left({\pi\over 2}\right)^{1/2}
\sum^\infty_{j=1/2}
\sqrt{2j+1}\left[(-1)^{j+1/2} - e^{-i\pi\nu}\right]
\left(\phi^{(1)}_{jm} - \phi^{(2)}_{jm}\right)
	\eqn\axvi  $$
where $\nu\equiv (j+1/2)/\ga$.
	The scattering amplitude $f(\gt)$ is determined form
$$ \psi_{\rm out} = f(\gt){e^{ikr}\over r}\chi
	\eqn\axvii  $$
where
$$ \chi \equiv {1\over [2(1-\cos\gt)]^{1/2}}\left(\matrix{\sin\gt\
e^{-i\gp}\cr 1-\cos\gt}\right)
	\eqn\axviii  $$
Letting $\gt \rightarrow \pi - \gt$,
$\gt$ now representing the scattering angle, we obtain
the scattering amplitude
$$ f(\gt) = {i\over 2k\cos\gt/2}\sum^\infty_{n=1}n\left(1 -
e^{-in\w\pi}\right)\left[P_{n-1}(\cos\gt) + P_n(\cos\gt)\right]
	\eqn\axix  $$
where $\w\equiv (1/\ga) - 1$, and $P_n$ is the Legendre polynomial.
The summation can be naively (in the
sense that we do not consider regularizing possibly ill-defined
sums) carried out to obtain
$$ f(\gt) = {i\over k}\gd(1-\cos\gt) + {\sin\w\pi/2\cos\gt/2
\over \sqrt{2}k(\cos\w\pi - \cos\gt)^{3/2} }
	\eqn\axx  $$
with the understanding that the $\gd$-function integrates to one
rather than a half.

	Note that $f(\gt)$ diverges at $\gt=0, \ \w\pi$. This is
similar to what happens in $2+1$ dimensions and is due to the
long-range nature of the ``interaction''---since the metric is
not asymptotically flat \refmark{\Deser}. Moreover, $f(\gt)$ does
not vanish in the ``no-interaction'' limit, $\w\rightarrow 0$. This
also happens in the case of non-relativistic Coulomb scattering,
where one is also
left with a $\gd$-function term in the no-interaction limit.

{\it 2. Fermion-Dyon Systems}

	First, we study fermion-dyon systems in the general spherically
symmetric background metric,
the Euclidean version of Eq.\general.
We consider a $SU(2)$ model with two massless left-handed
doublets. The action is given by $S=S_A + S_{A,\psi}$ with
$$\eqalign{S_A &= \int\sqrt{g}d^4x\left[{1\over 2e^2}{\rm tr}F^2_{\mu\nu}
- M_{mon} \right]   \cr
S_{A,\psi} &= i\sum^2_{s=1}
\int\sqrt{g}d^4x\ \bar\psi_L^{(s)}\gg^\mu(\p_\mu - \gG_\mu + A_\mu)\psi_L^{(s)}
}
	\eqn\si  $$
where $s = 1 , 2$ is the flavor index, $\gG_\mu$ is
the spin connection, and $M_{mon}$ is a constant.
We have neglected the Higgs field contribution to the action.
Spherically symmetric $SU(2)$ gauge fields are taken to be
$$\eqalign{A_0(r,t) &= -i\bgt\cdot\bn a_0(r,t)  \cr
A_i(r,t) &= -i\bgt\cdot\bn n_i{a_1(r,t)\over \ga(r)\gb(r)}
-i(\bn\times \bgt)_i{1-F(r)\over 2R(r)}   }
	\eqn\six    $$
In such a model,
the singlet chiral number currents are anomalous
$$\p_\mu\left(\sqrt{g}J^{\mu s}\right)\equiv\p_\mu\left(
\sqrt{g}\bar\psi_L^{(s)}\gg^\mu\psi_L^{(s)}\right) = -{1\over 16\pi^2}
\ge^{\mu\nu\gl\rho}{\rm tr}F_{\mu\nu}F_{\gl\rho}
\ ,\qquad s = 1 , 2
	\eqn\ssiv   $$
Therefore the gauge invariant chiral fermion number operators
$$ N^{(s)} \equiv \int d^3x \sqrt{g}J^{0s}\ ,\qquad s = 1 , 2
	\eqn\ssv   $$
are not conserved.

\def\D{{\cal D}}
\def\k{\kappa}

	We will use the representation of the flat space $\gg$-matrices
$$ \gg^0 =\left(\matrix{0&-i\cr i&0}\right) \ , \qquad
\bgg = \left(\matrix{0&\bgs\cr \bgs&0}\right ) \ , \qquad
\gg^5 = \left(\matrix{1&0 \cr 0&-1}\right )
	\eqn\ssiii    $$
For the two component, left-handed spinors $\psi^{(s)}_-$,
the Dirac equation reduces to
$$\eqalign{0 &=
\left[{1\over \gb}\p_t - \bgs\cdot\bp^{(\ga,R)} -
i{1-\ga R'\over R}\bgs\cdot
\bn + i\ga{\gb'\over 2\gb}\bgs\cdot\bn \right.\cr
&\qquad\left.-i{1\over \gb} a_0
\bgt\cdot\bn + {1\over \gb} a_1\bgs\cdot\bn\ \bgt\cdot\bn + {1-F(r)\over
2R}(\bgs\times\bn)\cdot\bgt \right]\psi^{(s)}_-   }
	\eqn\ssix    $$
where $\bgs\cdot\bp^{(\ga,R)}$ is given by Eq.\sxii\
with $\bgs$ replacing $\bgg$.
Spherically symmetric $SU(2)$ isospinor fermion fields are
of the form \Ref\Jackiw{ R.Jackiw
and C.Rebbi, Phys.Rev.D{\bf 13}, 3398 (1976).}
$$ \psi^{(s)}_-(r,t)_{in} =
{1\over \sqrt{8\pi\gb R^2}}\left[\chi^{(s)}_1(r,t)-i\chi^{(s)}_2(r,t)\bgs
\cdot\bn\right]_{im} (i\tau^2)_{mn}
	\eqn\ssvii   $$
where $i$ is the spin index and $n$ the isospin index. Introducing
$$\chi^{(s)} \equiv \left(\matrix{\chi^{(s)}_1\cr\chi^{(s)}_2}\right)
	\eqn\ssvi   $$
one finds that, in the $S$-wave approximation,
the currents $J^{\mu s}$ have the following reductions
$$J^{ts} = -{1\over 4\pi\gb^2R^2}\bar\chi^{(s)}\tgg^0\chi^{(s)} \ ,
\qquad J^{rs} = -{\ga\over 4\pi\gb R^2}\bar\chi^{(s)}\tgg^1\chi^{(s)}
	\eqn\current  $$
The original model can be reduced to the two dimensional model
described by the action $S = S_A + S_{A,\psi}$ with
$$\eqalign{ S_A &= {4\pi\over e^2}\int dxdt\left[(\p_xa_0-\p_ta_1)^2
{R^2(r)\over \gb^2(r)} \right]  \cr
S_{A,\psi} &= \sum_{s=1}^2\int
dxdt\ \bar\chi^{(s)}\tgg^\ga(i\p_\ga-\tgg^5a_\ga)\chi^{(s)}
\equiv i\sum_{s=1}^2\int dxdt\ \chi^{(s)\dagger}\D\ \chi^{(s)} }
	\eqn\ssxi   $$
where $\tgg^0=\gs^3 , \tgg^1=\gs^1 , \tgg^5=-i\tgg^0\tgg^1=\gs^2$ and
$s$ is the flavor index.
We have taken the zero magnetic monopole size limit and introduced
$$ dr=\ga(r)\gb(r) dx
	\eqn\eq   $$

	We see that the fermion action can always be reduced to a
two-dimensional, flat space form. This remains true for massive
fermions. Thus one can always integrate out
the fermion fields (for massless fermions) or bosonize the action.
These can be used to further understand systems with fermions and
magnetically charged black holes. Such problems were recently
introduced and studied in \Ref\Frank{C.Holzhey
and F.Wilczek, Nucl.Phys.B{\bf 380}, 447 (1992); \nextline
M.G.Alford and A.Strominger, Phys.Rev.Lett.{\bf 69}, 563 (1992).}.

	Next we specialize to the global monopole background
metric and calculate the density of the fermion condensate,
$<\psi^{(1)c\dagger}_-\psi^{(2)}_->^{mon}$,
where $1, 2$ are flavor indices and $\psi^{(s)c}_-$ is the fermion
number conjugation of $\psi^{(s)}_-$
$$ (\psi^{(s)}_-)^c_{in} \equiv (i\gs^2)_{ij}(i\tau^2)_{nm}
(\psi^{(s)}_-)^*_{jm}
	\eqn\sfvi  $$
One can verify that $\psi^{(s)c}_-$ satisfies the Dirac equation, Eq.\ssix,
provided $\psi^{(s)}_-$ satisfies the Dirac equation. Using the {\it Ansatz}
Eq.\ssvii\ we obtain
$$ \psi^{(1)c\dagger}_-\psi^{(2)}_- = {1\over 4\pi r^2}
(\chi^{(1)}_1\chi^{(2)}_1 + \chi^{(1)}_2\chi^{(2)}_2)
\equiv {1\over 4\pi r^2}f(x,t)
	\eqn\sfvii   $$
To calculate $<f(x,t)>^{mon}$, one uses the cluster property
$$ \lim_{|t_1-t_2|\rightarrow \infty}<f(x_1t_1)f^\dagger(x_2t_2)>^{mon}
= <f(x_1t_1)>^{mon}<f^\dagger(x_2t_2)>^{mon}
	\eqn\sfviii  $$
The remaining computation closely follows that of Callan and
Rubakov \Ref\Callan{C.G.Callan, Jr., Phys.Rev.D{\bf 25}, 2141 (1982);
V.A.Rubakov, Nucl.Phys. B{\bf 203}, 311 (1982).}.
It is to some extent similar to that
used in studying the massless Schwinger model
\Ref\Schwinger{ N.K.Nielsen
and B.Schroer, Nucl.Phys. B{\bf 120}, 62 (1977);
	 J.Schwinger, Phys.Rev. {\bf 128}, 2425 (1962).}.
	The first step is to find the exact fermion propagator,
which satisfies
\def\s{\gs^2}
\def\pt{\p_t}
\def\px{\p_x}
$$[\pt + i\s a_0 + i\s(\px + i\s a_1)]G(xt,x't')
	= \gd(x-x')\gd(t-t')
	\eqn\ssxii  $$
and the ``bag'' boundary condition \refmark{\Callan}
$$ (1 - \gs^3)G(0t,x't') = 0
	\eqn\ssxiii  $$
It can be solved by the {\it Ansatz} \refmark{\Schwinger}
$$ G(xt,x't') = \exp[\phi(x,t)]G_0(xt,x't')\exp[-\phi(x',t')]
	\eqn\ssxiv  $$
where $G_0$ is the free fermion propagator
satisfying the bag boundary condition, and is given by
\def\dm{\gD(x-x',t-t')}
\def\ddp{\gD(x+x',t-t')}
$$ G_0(xt,x't') = (\pt -i\s\px)[\dm + \ddp\gs^3]
	\eqn\eq $$
and $\phi$ is found to be
$$ \phi(x,t) = \zeta(x,t) -i\s\gg(x,t)
	\eqn\ssxviii  $$
with
$$\eqalign{\zeta(x,t) &= \pt\int dx'dt'[\dm + \ddp]a_1(x',t') \cr
&\qquad -\px\int dx'dt' [\dm - \ddp]a_0(x',t')  \cr
\gg(x,t) &= \px\int dx'dt'[\dm + \ddp]a_1(x',t') \cr
&\qquad +\pt\int dx'dt' [\dm - \ddp]a_0(x',t')  }
	\eqn\ssxix  $$
where $\gD$ is the free scalar propagator
$$ \gD(x,t) = {1\over 4\pi}\ln\mu^2(x^2 + t^2)
	\eqn\ssxx  $$
The choice of the boundary terms in Eq.\ssxix\ is to ensure that
$ \px\zeta(0,t) = 0$ and $\gg(0,t) = 0$,
so that the boundary condition Eq.\ssxiii\ can be satisfied.

	Under a gauge transformation
$ a_0\rightarrow a_0 - \pt\gb,\ a_1\rightarrow a_1 -\px\gb$,
one can verify that
$ \zeta\rightarrow \zeta, \ \gg\rightarrow \gg - \gb$.
Therefore the boundary condition does not change the gauge transformation
properties of the exact fermion propagator.
The gauge invariant expression for
${\rm tr}G(\xi,\xi)$ is
$$ {\rm tr}G(\xi,\xi) = {\rm tr}\left.\left[G(\xi,\xi')\exp\left(
i\s\int^\xi_{\xi'}d\tilde\xi^\ga a_\ga(\tilde\xi)\right)\right]
\right|_{\xi'\rightarrow\xi} 	= {1\over \pi}\pt\zeta
	\eqn\ssxxvi  $$
where we have introduced $(\xi^0,\xi^1)\equiv (t,x)$,
and the approach of $\xi'$ to $\xi$ is performed
from a spatial direction.

	One may calculate the induced vacuum currents
$$\eqalign{j^{\ga s} &= \left.<\bar\chi^{(s)}\tgg^\ga\chi^{(s)}>\right|_A
= {\rm tr}\left[\tgg^0\tgg^\ga G(\xi,\xi)\right]
= {1\over \pi}(\pt\zeta , \px\zeta)  \cr
j^{\ga s}_5 &= \left.<\bar\chi^{(s)}\tgg^5\tgg^\ga\chi^{(s)}>\right|_A
= {\rm tr}\left[\tgg^0\tgg^5\tgg^\ga G(\xi,\xi)\right]
= {i\over \pi}(\px\zeta , -\pt\zeta)  }
	\eqn\ssxxviii  $$
Therefore the axial-vector current is conserved,
$\p_\ga j^{\ga s}_5 = 0 $,
while the vector current possesses an anomaly
$$ \p_\ga j^{\ga s} = {1\over \pi} (\pt^2 + \px^2)\zeta
= -{1\over \pi}(\px a_0 - \pt a_1)
	\eqn\ssxxx  $$
which agrees with Eqs.\ssiv\ and \current.

\def\O{{\cal O}}
	The determinant
of an operator $\O$, depending on some parameter $\gl$,
may be computed from the Green's function $G(\xi,\xi')$ associated
with $\O$ as follows
$${\p\over\p\gl}\ln{\rm Det}\O = {\rm tr}{\p\O\over\p\gl}
\int d\xi G(\xi,\xi)
	\eqn\ssxxxv   $$
Using this, we obtain the fermion determinant
$$ \ln{\rm Det}\D = {1\over 2\pi}\int dxdt\ \zeta\square\zeta
 = \ln{\rm Det}i\D
	\eqn\ssxxxvi  $$
where $\square\equiv (\pt^2 + \px^2)$.
The effective action, obtained by integrating out the
fermion fields, is thus
$$ S_{eff}(\zeta) = S_A(\zeta) - 2\ln{\rm Det}i\D
= {1\over 2}\int dxdt\ \zeta(x,t)L_{xt}\zeta(x,t)
	\eqn\ssxxxvii  $$
where
$$ L_{xt}\equiv {2\over \pi}\left[-\square + {4\pi^2\ga^2\over
e^2}\square\ x^2\square\right]
	\eqn\ssxxxviii  $$
\def\P{{\cal P}}
The Green's function associated with $L_{xt}$, satisfying
$ \px\P(0t,x't') = 0 $, is given by
$$\P(xt,x't') = {\pi\over 2}\left[D_{e^2/4\pi^2\ga^2}(xt,x't')
- \dm -\ddp\right]
	\eqn\ssxxxx  $$
where $D_\k$ satisfies \refmark{\Callan}
$$ \left(\p_t^2 + \p_x^2 - {\k\over x^2}\right)D_\k(xt,x't')
	=\gd(x-x')\gd(t-t')
	\eqn\exact  $$
with $\k = e^2/4\pi^2\ga^2$. Eq.\exact\
can be solved by a Fourier transform with respect to
time. One finds
$$ \left(-\w^2 + \p_x^2 -{\k\over x^2}\right)\widetilde D_\k(x,x',\w)
	= {1\over 2\pi}\gd(x-x')
	\eqn\sfii   $$
Hence
$$ \widetilde D_\k(x,x',\w) = \cases{ -{\sqrt{xx'}\over 2\pi}I_\nu(|\w|
x)K_\nu(|\w| x')\ , &$\qquad x < x' $ \cr
 -{\sqrt{xx'}\over 2\pi}I_\nu(|\w|
x')K_\nu(|\w| x) \ , &$\qquad x > x'$ }
	\eqn\sfiii  $$
where $\nu\equiv \sqrt{\k + 1/4}$ and $I , K$ are modified Bessel
functions. One can then Fourier transform
back \Ref\table{I.S.Gradshteyn and I.M.Ryzhik, {\it Table of Integrals,
Series, and Products}, (Academic, New York, 1980) page 732.}
to obtain
$$ D_\k(xt,x't') = -{1\over 2\pi}Q_{\nu-1/2}\left[1+{(x-x')^2+
(t-t')^2\over 2xx'}\right]
	\eqn\sfiv  $$
where $Q$ is the Legendre function.

	The fermionic Green's functions in the presence of the
magnetic monopole are given by
$$\eqalign{ &<\chi(x_1t_1)\cdots\chi(x_Nt_N)
\chi^\dagger(x'_1t'_1)\cdots\chi^\dagger(x'_Nt'_N)>^{mon} \cr
&\qquad =\int
D\zeta e^{-S_{eff}(\zeta)}\left. <\chi(x_1t_1)\cdots\chi(x_Nt_N)
\chi^\dagger(x'_1t'_1)\cdots\chi^\dagger(x'_Nt'_N)>\right|_A  }
	\eqn\sfv   $$
To calculate the density of condensate,
one obtains $<f(x_1t_1)f^\dagger(x_2t_2)>^{mon}$ from
$$\eqalign{<f(x_1t_1)&f^\dagger(x_2t_2)>\Big|_A =
\left.<\chi^{(1)}_\ga(x_1t_1)\chi^{(2)}_\ga(x_1t_1)
\chi^{(1)\dagger}_\gb(x_2t_2)\chi^{(2)\dagger}_\gb(x_2t_2)>
\right|_A   \cr
&= {\rm tr}\left[G_0(x_1t_1,x_2t_2)G^{\rm T}_0(x_1t_1,x_2t_2)\right]
\exp\left[2\zeta(x_1,t_1)-2\zeta(x_2,t_2)\right]  }
	\eqn\sfix   $$
where we have used Wick's theorem and Eqs.\ssxiv, \ssxviii.
Therefore
$$ \eqalign{<f(x_1t_1)f^\dagger(x_2t_2)>^{mon} &=
 {\rm tr}\left[G_0(x_1t_1,x_2t_2)G^{\rm T}_0(x_1t_1,x_2t_2)\right] \cr
\times\int D\zeta &\exp\left[-S_{eff}(\zeta) +
2\zeta(x_1,t_1)-2\zeta(x_2,t_2)\right]  \cr
&= {\rm tr}\left[G_0(x_1t_1,x_2t_2)G^{\rm T}_0(x_1t_1,x_2t_2)\right] \cr
\times\exp [ -4&\P(x_1t_1,x_2t_2) + 2\P(x_1t_1,x_1t_1)
+ 2\P(x_2t_2,x_2t_2) ]  \cr
&= {1\over 4\pi^2 x_1x_2}\left[1 + O\left({e^2\over \ga^2}\right)
\right]   }
	\eqn\sfxii   $$
where we have used the saddle-point method and
taken the limit $|t_1-t_2|\rightarrow \infty$ on the
last line, and used properties of the Legendre function
\refmark{\Callan}.
According to the cluster property, Eq.\sfxii\ implies
$$ <f(x,t)>^{mon} = {e^{i\gt}\over 2\pi x}\left[1 + O\left({e^2
\over \ga^2}\right)\right]
	\eqn\sfxiii   $$
where $\gt$ is an arbitrary phase. Recall that $x\equiv r/\ga$,
we have thus found that the density of the fermion number
condensate is $\ga$ times the flat space value. Therefore the amount
of condensate from radius $r$ to $r+dr$ is the same as in flat space.

\ack

	I thank Professor Erick Weinberg for helpful discussions
and valuable comments on the manuscript, Professor Roman Jackiw for
a critical reading of the paper.

\refout

\end